\documentclass[aps,preprint,showpacs,showkeys,sort]{revtex4}
\usepackage{amsfonts}
\usepackage{amsmath}
\usepackage{amssymb}
\usepackage{graphicx}

\begin{document}

\preprint{HEP/123-qed}
\title{Information metric from Riemannian superspaces.}
\author{Diego Julio Cirilo-Lombardo}
\affiliation{International Institute of Physics, Natal, RN, Brazil}
\affiliation{Bogoliubov Laboratory of Theoretical Physics Joint Institute for Nuclear
Research, 141980, Dubna, Russian Federation}
\author{Victor I. Afonso}
\affiliation{Unidade Acad\^{e}mica de F\'{\i}sica, Universidade Federal de Campina
Grande, PB, Brasil}
\pacs{PACS number}

\begin{abstract}
The Fisher's information metric is introduced in order to find the real
meaning of the probability distribution in classical and quantum systems
described by Riemaniann non-degenerated superspaces. In particular, the
physical r\^{o}le played by the coefficients $\mathbf{a}$ and $\mathbf{a}^\ast$ of the pure fermionic part of
a genuine emergent metric solution, obtained in previous work \cite{diegoSuper} is explored. 
To this end, two characteristic viable distribution functions 
are used as input in the Fisher definition: 
first, a Lagrangian generalization of the Hitchin Yang-Mills prescription and,
second, the probability current associated to the emergent non-degenerate superspace geometry.
Explicitly, we have found that the metric solution of the superspace
allows establish a connexion between the Fisher metric
and its quantum counterpart, corroborating early conjectures by Caianiello {\em et al.}
This quantum mechanical extension of the Fisher metric is described by
the $CP^1$ structure of the Fubini–Study metric, with coordinates $\mathbf{a}$ and $\mathbf{a}^\ast$.
\end{abstract}

\maketitle
\tableofcontents

\section{Introduction}

The problem of giving an unambiguous quantum mechanical description of a
particle in a general spacetime has been repeatedly investigated. 
The introduction of supersymmetry provided a new approach to this question,
however, some important aspects concerning the physical observables remain
not completely understood, classically and quantically speaking.

The superspace concept, on the other hand, simplify considerably the link
between ordinary relativistic systems and `supersystems', 
extending the standard (bosonic) spacetime by means of a general (super)group manifold,
equipped with also fermionic (odd) coordinates.

In a previous work \cite{diegoSuper} we introduced, besides other supersymmetric
quantum systems of physical interest, a particular $N=1$ superspace \cite{diegoFund}.
That was made with the aim of studying a superworld-line quantum particle 
(analogously to the relativistic case) and its relation with SUGRA theories \cite{va, diegoFund, diegoPLB}. 
The main feature of this superspace is that the supermetric,
which is the basic ingredient of a Volkov-Pashnev particle action \cite{pash, va},
 is \emph{invertible and non-degenerate}, 
 that is, of  $G4$ type in the Casalbuoni's classification \cite{casal}. 
As shown in \cite{diegoFund, diegoPLB}, the non-degeneracy of the supermetrics 
(and therefore of the corresponding superspaces) 
leads to important consequences in the description of physical systems.
In particular, notorious geometrical and topological effects on the quantum states,
namely, \textit{consistent mechanisms of localization and confinement}, 
due purely to the geometrical character of the Lagrangian. 
Also \textit{an alternative to the Randall-Sundrum (RS) model without extra bosonic coordinates},
 can be consistently formulated in terms of such non degenerated superspace approach, 
 eliminating the problems that the RS-like models present at the quantum level \cite{diegoSuper, diegoPLB}.

Given the importance of the non degeneracy of the supermetrics in the
formulation of physical theories, in the present article we analyze further
the super-line element introduced in \cite{pash, diegoFund}, focusing on a probabilistic context.
To this specific end, the Fisher metric \cite{fisher} (Fisher-Rao in the quantum sense \cite{rao}) 
have been considered in several works in order to provide a geometrical interpretation of the statistical measures. 
Fisher's information measure (FIM) was advanced already in the 1920's decade, 
well before the advent of Information Theory (IT). 
Much interesting work has been devoted to the physical applications of FIM in recent times 
(see, for instance, \cite{frieden} and references therein). 
In \cite{ponjas}, a generalization of the Yang-Mills Hitchin proposal was made
suggesting an indentification of the Lagrangian density with the Fisher probability distribution ($P\left( \theta \right)$). 
However, this idea was explored from a variational point of view, in previous work by Plastino {\it et al.} \cite{plastino, fp}
This proposal brought a contribution to the line of works 
looking for a connexion between the spacetime geometry and quantum field theories.

In the last decades it has been claimed that the above expectation is partially realized in the AdS/CFT 
(anti-de Sitter/Conformal Field Theory) correspondence \cite{maldacena}, 
which asserts that the equivalence of a gravitational theory (i.e., the geometry of spacetime) 
and a conformal quantum field theory at the boundary of spacetime certainly exists.

The organization of this paper is as follows: in Section \ref{sec:II}, the Fisher metric and a new
generalization of the Hitchin proposal are introduced. 
Section \ref{sec:III}  presents global aspects of the $N=1$ non-degenerate superspace solution of reference \cite{diegoSuper}.
In Section \ref{sec:susybreak} we analyze the `bosonic' ($B_0$) part of our Fisher supermetric by `turning off' all the fermions.
Section \ref{sec:further} presents further discussions on our results, in connexion with a quantum extension of the Fisher metric.
Finally, Section \ref{sec:conclusions} is devoted to collect the main results and our concluding remarks.

\section{Fisher's metric and Hitchin's prescription}\label{sec:II}

In general, the Fisher information metric \footnote{We prefer the term `information metric' rather than `emergent metric' because an `emergent metric' appears as a quantum solution for the physical spacetime geometry while the Fisher information metric is related to statistical parameters and is, therefore,a probabilistic concept}  
(more precisely, the Fisher-Rao metric) is a Riemannian metric for the manifold of the parameters of probability distributions.
The Rao distance (geodesic distance in the parameter manifold) 
provides a measure of the difference between distinct distributions.
In the thermodynamic context, the Fisher information metric is directly related to the rate of change in the corresponding order parameters 
and can be used as an information-geometric complexity measure for classifying phase transitions,
{\em e.g.}, the scalar curvature of the thermodynamic metric tensor diverges at (and only at) 
 a phase transition point (this issue will be analyzed in future work).
In particular, such relations identify second-order phase transitions via divergences of individual matrix elements. 

The Fisher-Rao information metric is given by \cite{fisher, rao}
\begin{equation}
G_{ab}(\theta)=\int{d^Dx \, P(x;\theta)}\,\partial_a\ln{P(x;\theta)}%
\,\partial_b\ln{P(x;\theta)}  \label{gfisher}
\end{equation}
where $x^\mu$ ($\mu, \nu=0,\ldots, D$) are the random variables and $\theta_a$ ($a, b= 1,\ldots, N $) are the parameters of the probability
distribution. Besides this, $P(x;\theta)$ must fulfil the normalization condition 
\begin{equation}
\int{d^Dx P(x;\theta)}=1  \label{norm}
\end{equation}

In his work \cite{hitchin}, Hitchin proposed the use of the the squared
field strength of Yang-Mills theory as a probability distribution.
A generalization of the Hitchin's proposal \cite{ponjas} -- see also \cite{plastino} --
 consist in identifying the probability distribution with the \emph{on-shell} Lagrangian
density of a field theory
\begin{equation}
P(x;\theta):=-\mathcal{L}(x;\theta)|_{solution}.
\end{equation}

In Ref. \cite{ponjas}, they consider the case in which the distribution
dependence on the variables takes the form $P(x,\theta)=P(x-\theta)$. 
Note that this is only possible in the cases in which the numbers of
the space-time dimensions $x$ and the parameters $\theta$ coincide, that is, 
in which $D=N$.

In the present work we will consider two different approaches to the
problem. First, we will follow the `generalized Hitchin prescription',
identifying our Lagrangian (calculated at the solution) with the probability
distribution. Then, we will introduce a new proposal: we will take the
state probability current of the emergent metric solution of the superspace (obtained explicitly in \cite{diegoSuper})
as beeing itself the probability distribution. The results of the two approaches will be
compared in order to infer the physical meaning of the $\mathbf{a}$ and $%
\mathbf{a}^\ast$ parameters appearing in the pure fermionic part of the
superspace metric.

\section{Emergent metric solution} \label{sec:III}

The model introduced in \cite{diegoSuper} describes a free particle in a
superspace with coordinates $z_{A}\equiv \left( x^{\mu },\theta _{\alpha },%
\overline{\theta }_{\overset{\cdot }{\alpha }}\right)$. 
The corresponding Lagrangian density is
\begin{equation}
\mathcal{L}=-m \sqrt{\omega^A\omega_A}=-m\sqrt{\overset{\circ}{\omega}_\mu%
\overset{\circ}{\omega}^{\,\mu} +{\mathbf{a}}\dot{\theta}^\alpha \dot{\theta}%
_\alpha- \mathbf{a}^{\ast}\dot{\bar{\theta}}^{\dot\alpha}\dot{\bar{\theta}}%
_{\dot\alpha}}.  \label{CSL}
\end{equation}
where $\overset{\circ }{\omega _{\mu }}=\overset{.}{x}_{\mu }-i(\overset{.}{%
\theta }\ \sigma _{\mu }\overline{\theta }-\theta \ \sigma _{\mu }\overset{.}%
{\overline{\theta }})$, and the dot indicates derivative with respect to the
evolution parameter $\tau $, as usual. In coordinates, the line element of the
superspace reads, 
\begin{equation}
ds^2=\dot{z}^{A}\dot{z}_{A} =\dot{x}^{\mu }\dot{x}_{\mu }-2i\dot{x}^{\mu }(%
\dot{\theta}\sigma _{\mu }\bar{\theta}-\theta \sigma _{\mu }\dot{\bar{\theta}%
})+\left( {\mathbf{a-}}\bar{\theta}^{\dot{\alpha}}\bar{\theta}_{\dot{\alpha}%
}\right) \dot{\theta}^{\alpha }\dot{\theta}_{\alpha }-\left( {\mathbf{a}}%
^{\ast }+\theta ^{\alpha }\theta _{\alpha }\right) \dot{\bar{\theta}}^{\dot{%
\alpha}}\dot{\bar{\theta}}_{\dot{\alpha}} 
\end{equation}

The `squared' solution with three compactified dimensions ($\lambda$ spin
fixed) is \cite{diegoSuper}  
\begin{equation}
g_{AB}(t)= e^{A(t)+\xi\varrho(t)}g_{AB}(0),  \label{CSsol}
\end{equation}
where the initial values of the metric components are given by 
\begin{equation}
g_{ab}(0)=\langle \psi (0)|\left( 
\begin{array}{c}
a \\ 
a^{\dagger }%
\end{array}%
\right) _{ab}|\psi (0)\rangle,
\end{equation}%
or, explicitly, 
\begin{eqnarray}
g_{\mu \nu }(0) =\eta _{\mu \nu } \,,\qquad g_{\mu \alpha }(0) =-i\sigma
_{\mu \alpha \dot{\alpha}}\bar{\theta}^{\dot{\alpha}} \,,\qquad g_{\mu \dot{%
\alpha}}(0) =-i\theta ^{\alpha }\sigma _{\mu \alpha \dot{\alpha}} \,, \label{gdiego}\\
g_{\alpha \beta }(0) =(a-\bar{\theta}^{\dot{\alpha}}\bar{\theta}_{\dot{\alpha%
}})\epsilon _{\alpha \beta } \,, \qquad g_{\dot{\alpha}\dot{\beta}}(0)
=-(a^{\ast }+\theta ^{\alpha }\theta_{\alpha })\epsilon _{\dot{\alpha}\dot{%
\beta}} \,.  \nonumber
\end{eqnarray}
It is worth mention here that these components were obtained in a simpler case in \cite{va}.

The bosonic and spinorial parts of the exponent in the superfield solution (\ref{CSsol}) are, respectively, 
\begin{eqnarray}
A(t) & = & -\left(\frac{m}{|{\mathbf{a}}|}\right)^{2} t^{2}+c_{1}t+c_{2}, \label{A}\\ 
\text{and} \qquad\qquad&&\nonumber\\
\xi\varrho(t) & = & \xi\left(\phi_{\alpha}(t)+\bar{\chi}_{\dot{\alpha}}(t)\right) \nonumber\\
& = & \theta^{\alpha}\left(\overset{\circ}{\phi}_{\alpha}\cos(\omega t/2) +\tfrac{2}{\omega} Z_{\alpha}\right)-\bar{\theta}^{\dot{\alpha}}
\left(-\overset{\circ}{\bar{\phi}}_{\dot{\alpha}}\sin\left(\omega t/2\right) -\tfrac{2}{\omega}\bar{Z}_{\dot{\alpha}}\right) \label{expo}\\ 
& = & \theta^{\alpha}\overset{\circ}{\phi}_{\alpha }\cos(\omega t/2) +\bar{\theta}^{\dot\alpha}
\overset{\circ }{\bar{\phi}}_{\dot\alpha}\sin(\omega t/2)+4|\mathbf{a}|Re(\theta Z),\nonumber
\end{eqnarray}
where $\overset{\circ }{\phi }_{\alpha },Z_{\alpha },\overline{Z}_{\overset{.%
}{\beta }}$ are constant spinors, 
$\omega \approx 1/|\mathbf{a}|$ and the constant $c_{1}$, 
due to the obvious physical reasons and the chirality restoration of the superfield solution \cite{diegoSuper,diegoFund},
should be taken purely imaginary.

\section{Fisher information metric from Riemannian superspaces}\label{sec:IV}

Fisher method considers a familiy of probability distributions, characterized by
certain number of parameters. The metric components are then defined by
considering derivatives in different `directions' in the parameters space.
That is, measuring `how distant' two distinct set of parameters put apart the corresponding probability distributions.

In the following we will calculate the Fisher information metric corresponding to
a generalized Hitchin `on-shell' Lagrangian prescription.
In our case, the parameters of interest in the metric solution (\ref{CSsol}) are $\mathbf{a}$ and 
$\mathbf{a}^\ast$, which could indicate the residual effects of supersymmetry 
given that they survive even when `turning off' all the fermionic fields.

\subsection{Generalized Hitchin prescription for the probability distribution} \label{ssec:hitch}

Following the generalized Hitchin prescription we identify the probability
ditribution with Lagrangian (\ref{CSL}), evaluated at solution (\ref{CSsol}), 
$g_{AB}(t)= e^{(A(t)+\xi\varrho(t))} g_{AB}(0)$. Thus, the
probability distribution density takes the form 
\begin{equation}
P\left(z^A,a,a^\ast\right):=-\mathcal{L}|_{g_{AB}(t)} =\, e^{\frac12 \left(
A(t)+\xi\varrho(t)\right)}\mathcal{L}_{0},
\end{equation}
where 
\begin{equation}
\mathcal{L}_{0}\equiv\mathcal{L}(g_{ab}(0))=m\sqrt{\overset{\circ }{\omega
^{\nu }}\overset{\circ }{\omega ^{\,\mu }}+{\mathbf{a}}\dot{\theta}^{\alpha }%
\dot{\theta}_{\alpha }-\mathbf{a}^{\ast }\dot{\bar{\theta}}^{\dot{\alpha}}%
\dot{\bar{\theta}}_{\dot{\alpha}}}.
\end{equation}

Let us now calculate the $\mathbf{a}$ and $\mathbf{a}^\ast$ derivatives of our probability distribution
\begin{eqnarray}
\frac{\partial P}{\partial \mathbf{a}} &=&\mathcal{L}_{0}\left[ \frac{1}{2}%
\frac{\partial (A(t)+\xi \varrho (t))}{\partial \mathbf{a}}+\frac{1}{%
\mathcal{L}_{0}}\frac{\partial \mathcal{L}_{0}}{\partial \mathbf{a}}\right]
\,e^{\frac{1}{2}(A(t)+\xi \varrho (t))}, \\
\notag \\
\frac{\partial P}{\partial \mathbf{a}^{\ast }} &=&\mathcal{L}_{0}\left[ 
\frac{1}{2}\frac{\partial (A(t)+\xi \varrho (t))}{\partial \mathbf{a}^{\ast }%
}+\frac{1}{\mathcal{L}_{0}}\frac{\partial \mathcal{L}_{0}}{\partial \mathbf{a%
}^{\ast }}\right] e^{\frac{1}{2}(A(t)+\xi \varrho (t))}.
\end{eqnarray}

The last terms in the squared brakets give 
\begin{equation}
\frac{1}{\mathcal{L}_{0}}\frac{\partial \mathcal{L}_{0}}{\partial \mathbf{a}}%
=\frac{m^2}{2\mathcal{L}_{0}^{2}}\,\dot{\theta}^{\alpha }\dot{\theta}%
_{\alpha }\qquad \text{and}\qquad \frac{1}{\mathcal{L}_{0}}\frac{\partial 
\mathcal{L}_{0}}{\partial \mathbf{a}^{\ast }}=-\frac{m^2}{2\mathcal{L}%
_{0}^{2}}\, \dot{\bar{\theta}}^{\dot{\alpha}}\dot{\bar{\theta}}_{\dot{\alpha}%
}
\end{equation}

From (\ref{expo}) we have 
\begin{eqnarray}
\Xi(t;|a|)&\equiv&\frac{\partial (A(t)+\xi \varrho (t))}{\partial |\mathbf{a}%
|}  \\
&=&\frac{2m^{2}}{|\mathbf{a}|^{3}}t^{2}+\frac{\omega ^{2}t}{2}\left( \theta
^{\alpha }\overset{\circ }{\phi }_{\alpha }\sin (\omega t/2)-\bar{\theta}^{%
\dot{\alpha}}\overset{\circ }{\bar{\phi}}_{\dot{\alpha}}\cos (\omega
t/2)\right) +4Re(\theta Z),\nonumber
\end{eqnarray}

Finally, putting $|\mathbf{a}|=\sqrt{\mathbf{a}\mathbf{a}^{\ast }}$ we have
that 
\begin{equation}
\frac{\partial |\mathbf{a}|}{\partial \mathbf{a}}=\frac{\mathbf{a}^{\ast }}{%
2|\mathbf{a}|}\qquad \text{and}\qquad \frac{\partial |\mathbf{a}|}{\partial 
\mathbf{a}^{\ast }}=\frac{\mathbf{a}}{2|\mathbf{a}|},  \label{da}
\end{equation}

We can now write down the Fisher's metric components 
\begin{eqnarray}
G_{aa}&=&\int {dx^{4}P^{-1}\left[ \frac{\partial P}{\partial \mathbf{a}}%
\right] ^{2}} =\frac{\mathcal{L}_{0}}{16}\int dt \left( \frac{\,\mathbf{a}%
^{\ast }}{|\mathbf{a}|}\,\Xi(t;|a|) +\frac{2 m^2}{\mathcal{L}_{0}^{2}}\dot{%
\theta}^{\alpha }\dot{\theta}_{\alpha}\right)^{2} \,e^{\frac{1}{2}(A(t)+\xi
\varrho(t))} \nonumber\\
G_{a^{\ast }a^{\ast }}\!\!&=&\!\int {dx^4 P^{-1}\left[ \frac{\partial P}{\partial 
\mathbf{a}^{\ast}}\right] ^{2}} =\frac{\mathcal{L}_{0}}{16}\int dt 
\left( \frac{\mathbf{a}}{|\mathbf{a}|}\,\Xi(t;|a|) -\frac{2 m^2}{\mathcal{L}%
_{0}^{2}}\dot{\bar{\theta}}^{\dot{\alpha}}\dot{\bar{\theta}}_{\dot{\alpha}%
}\right) ^{2} \,e^{\frac{1}{2}(A(t)+\xi \varrho(t))} \\
G_{aa^\ast}\!&=&\!g_{a^\ast a}=\int{dx^4 P^{-1} \frac{\partial P}{\partial 
\mathbf{a}}\frac{\partial P}{\partial \mathbf{a^\ast}}} 
=\frac{\mathcal{L}_{0}}{16} \int dt \left(\Xi(t;|a|)^2 +4\frac{%
m^4}{\mathcal{L}_0^4} \,\dot\theta^\alpha \dot\theta_\alpha \,\dot{\bar\theta%
}^{\dot\alpha} \dot{\bar\theta}_{\dot\alpha} \right) \,e^{\frac{1}{2}%
(A(t)+\xi \varrho (t))}\nonumber
\end{eqnarray}

\subsection{State probability current as distribution}\label{ssec:j0}

Our second approach consist in identifying the state probability density 
(zero component of the probability current) of the solution (\ref{CSsol})
as the probability density itself.

The zero component of the probablity current can be obtained by makeing
\begin{equation}
j_0(t) = 2 E^2  g_{ab}(t)  g^{ab}(t) .
\end{equation}
Then, putting  $\mathcal{K}_0 \equiv 32 E^2 |\alpha|^2$, we have
\begin{eqnarray}
j_0(t)&=& \frac{1}{16}\mathcal{K}_0 e^{-2\left( \frac{m}{|{\mathbf{a}}|}%
\right)^{2}t^{2}+ 2c_{2}+ 2\xi\varrho(t)},  \label{j0}\\
\text{and}\hspace{3cm}&& \nonumber\\
\frac{\partial j_0}{\partial |{\mathbf{a}}|}&=& \frac{1}{8}\mathcal{K}%
_0\,\Xi(t;|a|) e^{-2\left( \frac{m}{|{\mathbf{a}}|}\right)^{2}t^{2}+ 2c_{2}+
2\xi\varrho(t)}.\label{dj0}
\end{eqnarray}

Therefore, taking $P \equiv j_0(t)$ and using again (\ref{da}) we get
\begin{eqnarray}
G_{aa}&=&\int dx^4 P^{-1} \left[\frac{\partial P}{\partial \mathbf{a}}\right]%
^2 =\left(\frac{\,\mathbf{a}^\ast}{|\mathbf{a}|} \right)^2 {\cal I}_{\Xi} \label{Gaa} \\
G_{a^\ast a^\ast}\!\!&=&\!\!\int dx^4 P^{-1} \left[\frac{\partial P}{\partial \mathbf{a^\ast}}\right]^2 
=\left(\frac{\mathbf{a}}{|\mathbf{a}|} \right)^2 {\cal I}_{\Xi}  \label{GaAstaAst} \\
G_{aa^\ast}\!\!&=&G_{a^\ast a}\!=\!\int{dx^4 P^{-1} \frac{\partial P}{\partial 
\mathbf{a}}\frac{\partial P}{\partial \mathbf{a^\ast}}} ={\cal I}_{\Xi},  \label{GaaAst}
\end{eqnarray}
where ${\cal I}_{\Xi}$ corresponds to the integral of the temporal part
\begin{equation}
{\cal I}_{\Xi} \equiv\frac{1}{16}\mathcal{K}_0\int dt \; \Xi(t;|a|)^2 e^{-2\left(\frac{m}{|{\mathbf{a}}|}%
\right)^{2}t^{2}+2c_{2}+2\xi\varrho(t)}.
\end{equation}

\section{The $B_0$ parte of the Fisher supermetric}\label{sec:susybreak}

\subsection{Generalized Hitchin prescription with zero fermions}

When putting all fermion to zero, the derivative of the time dependent
exponential and the $\mathcal{L}_{0}$ initial value `on-shell' Lagrangian reduce,
respectively, to 
\begin{equation}
\Xi(t;|a|)|_{\theta=\chi=0}=\frac{2m^{2}}{|\mathbf{a}|^{3}}t^{2} \qquad 
\text{and}\qquad \mathcal{L}%
_{0}|_{\theta=\chi=0}=m\sqrt{\dot x^{\mu}\dot x_{\mu}}=m,  \label{Xiboson}
\end{equation}
where the last equality makes explicit the relativistic constraint $v^2 \equiv \dot x^{\mu}\dot x_{\mu}=1$.

In that case, and writing the complex parameters as $\mathbf{a}=|\mathbf{a}|e^{i\phi}$, the metric components take the simple form 
\begin{equation}
G_{ab}={\cal I}(m, c_1, c_2, |\mathbf{a}|) \left(
\begin{array}{cc}
e^{-i 2\phi} &1\\
1  & e^{i 2\phi}
\end{array}\right),\label{Gab}
\end{equation}
where indices $a,b$ take values in $\{\mathbf{a},\mathbf{a}^\ast\}$, 
and the prefactor is the integral of the time varying factor, that can be easily performed to obtain
\begin{eqnarray}
{\cal I}(m, c_1, c_2, |\mathbf{a}|) &\equiv& \frac{1}{4}\left(\frac{m^5}{|\mathbf{a}|^6}\right)
\int dt  \; t^4 \,e^{-\left( \frac{m}{|{\mathbf{a}}|}\right)^{2}t^{2}+c_{1}t+c_{2}} \label{I0f} \\
&=&\frac{\sqrt{\pi}}{64} 
\left[\left(\frac{c_1}{m}\right)^4 |\mathbf{a}|^3+ 12 \left(\frac{c_1}{m}\right)^2 |\mathbf{a}|+ 12 |\mathbf{a}|^{-1}\right]
e^{\frac{1}{4}\left(\frac{c_1}{m}\right)^2 |\mathbf{a}|^2 + c_2}  \nonumber
\end{eqnarray}

Note that the metric (\ref{Gab}) above can also be put in terms of Pauli's matrices as 
\begin{equation}
G_{ab}~=~{\cal I}(m, c_1, c_2, |\mathbf{a}|) \left(\mathbb{I}_2 cos(\phi) - i\sin(\phi) \sigma_z +\sigma_x \right)
\end{equation}

\subsection{State probability current with zero fermions}

Turning off all the fermions in solution (\ref{j0}), the zero component of
the probability distribution current reduces to 
\begin{equation}
j_{0}|_{\theta=\chi=0}= \frac{1}{16}\mathcal{K}_0 e^{ -2\frac{m^{2}}{|{\mathbf{a}}|^{2}} t^{2}+2c_{2}},
\end{equation}
and its derivative to
\begin{equation}
\left.\frac{\partial j_0}{\partial |{\mathbf{a}}|}\right|_{\theta=\chi=0}= 
\frac{1}{4} \frac{m^{2}}{|\mathbf{a}|^{3}}\mathcal{K}_0 \, t^{2} 
e^{-2\frac{m^{2}}{|{\mathbf{a}}|^{2}} t^{2}+2c_{2}}.
\end{equation}

Therefore, the metric components in this case take the simple form 
\begin{equation}
G_{ab}={\cal J}(m,E,|\alpha|,c_2,|\mathbf{a}|) 
\left(\begin{array}{cc}
e^{-i 2\phi}&1\\ 1  & e^{i 2\phi}
\end{array}\right).
\end{equation}
Again, $a,b$ taking values in $\{\mathbf{a},\mathbf{a}^\ast\}$, and
now the prefactor is obtained performing the integral
\begin{eqnarray}
{\cal J}(m,E,|\alpha|,c_2,|\mathbf{a}|) &\equiv& \frac{1}{4} \left(\frac{\mathcal{K}_0 m^4}{|\mathbf{a}|^6} \right)
\int dt \, t^4 \, e^{-2\left( \frac{m}{|{\mathbf{a}}|}\right)^{2}t^{2}+ 2c_{2}} \label{J0f}\\
&=& \frac{3\sqrt{2\pi}}{4}\frac{E^2 |\alpha|^2}{m} e^{2c_2}|\mathbf{a}|^{-1},\nonumber
\end{eqnarray}
where in last equality we have made explicit the value of $\mathcal{K}_0$.

Note that this last expression, in contrast with eq. (\ref{I0f}), 
presents only the $|{\mathbf{a}}|^{-1}$ singular term.
This is precisely due to the lack of the `free wave' (linear in $t$) term in the exponential factor,
which leads to a complete departure of the Gaussian behaviour shown by (\ref{I0f}),
remaining in common just the singular term.

\section{Information metric and geometrical Lagrangians: further discussion} \label{sec:further}
Note that the Fisher metric can be rewritten in the form
\begin{eqnarray}
G_{ab}(\theta ) &=&\int {d^{D}x\,P(x;\theta )}\,\partial _{a}\ln {P(x;\theta
)}\,\partial _{b}\ln {P(x;\theta )} \nonumber\\
&=&4\int {d^{D}x\,}\,\partial _{a}{P}^{1/2}{(x;\theta )}\,\partial _{b}{P}%
^{1/2}{(x;\theta )}. \label{GabPhalf}
\end{eqnarray}%
The appearence of the square root of the probability density $P$
above, naturally leads to the identification 
\begin{equation}
{P}^{1/2}\equiv \mathcal{L}_{g}\quad\Rightarrow\quad P=ds^{2}
\end{equation}%

Under this point of view, the $P$ function is related to the line element that define the
geometrical (superspace in our case) Lagrangian of the theory.
Therefore, this is a first approach to connect the two ``distances'': the Rao distance in the probability parameters manifold,
and the geometric space-time distance.

\subsection{$P$ as the current of probability: the quantum correspondence} \label{sec:geom}

Consider a Hilbert space $\mathcal{H}$ with a symmetric inner product $%
G_{ij}$. For instance, we can have in mind the case  $\mathcal{H}=\mathbf{L}_{2}(\mathcal{R})$,   
where $\mathcal{R}=\mathbb{R}^{2m}$ (e.g.: the phase space of a classical dynamical system, 
the configuration space spin systems, etc.) 
The equation (\ref{GabPhalf}), as was first observed by Caianiello et al. \cite{caia}, 
puts in evidence the clear possibility of mapping the probability density function ${P(x;\theta )}$ 
on $\mathcal{R}$ to $\mathcal{H}$ by forming the square-root.

As we propose in the present paper, using directly (and precisely) the probability current $j_{0}$,
of the metric state solution $g_{AB}(t)$, as the probability density ${P(x;\theta)}$, 
\begin{equation}
j_{0}=\frac{1}{16}\mathcal{K}_{0}e^{-2\left( \frac{m}{|{\mathbf{a}}|}\right)
^{2}t^{2}+2c_{2}+2\xi \varrho (t)}\equiv {P(x;\theta )},
\end{equation}
leads to an identification (intuited by Caianiello in \cite{caia}) that can be immediately implemented and effectively realized.
The metric components take then the form
\begin{eqnarray}
G_{ab}(\theta ) &=&4\int {d^{D}x\,}\,\partial_{a}\, {P}^{1/2}{(x;\theta )}%
\,\partial_{b}\, {P}^{1/2}{(x;\theta )} \nonumber\\
&\,\rightarrow\,& 4 \int d^{D}x\,\partial_{a}g_{AB}(x;\mathbf{a,a}^\ast)\,\partial_b\, g^{AB}(x;\mathbf{a,a}^\ast) \\
&=&4\int d^{D}x \,\partial_{a}\, g_{AB}(x;\mathbf{a,a}^\ast)\,
\partial_{b}\, g_{CD}(x;\mathbf{a,a}^\ast) \eta^{(AB)(CD)}. \nonumber
\end{eqnarray}
Then, the quantum `crossover' is
\begin{eqnarray}
G_{ab}(\theta ) &=&4 \,\partial_{a}\,g_{AB}(x;\mathbf{a,a}^\ast)\,
\partial_{b}\,g_{CD}(x;\mathbf{a,a}^\ast)\eta^{(AB)(CD)} \label{cross}\\
&\equiv & \left\langle \,\partial_{a}\, g_{AB}(x;\mathbf{a,a}^\ast)\,
\partial_{b}\, g_{CD}(x;\mathbf{a,a}^\ast)\right\rangle .\nonumber
\end{eqnarray}

The resemblance with a Sigma model is quite evident. 
Even more, the `natural' geometrical normalization condition 
$\;g_{AB}(x;\mathbf{a,a}^\ast)\,g^{AB}(x;\mathbf{a,a}^\ast)=D\;$
(regarding the genuine emergent origin of the spin 2 state $g_{AB}(x;\mathbf{a,a}^\ast)$), 
suggests a pseudospherical constraint in $\mathcal{H}$. 
However, the above relation (\ref{cross}) is not the more general possibility 
of quantum extension of the Fisher's metric. 
We can also introduce the following Hermitian metric tensor 
\begin{eqnarray}
\widetilde{G}_{ab}(\theta )&=&\left\langle\,\partial_{a}\,g_{AB}(x;\mathbf{a,a}^\ast)\,
\partial _{b}\,g_{CD}(x;\mathbf{a,a}^\ast)\,\right\rangle \\
&&-\left\langle \,\partial_{a}\,g_{AB}(x;\mathbf{a,a}^\ast)\,g_{CD}(x;\mathbf{a,a}^\ast)\,\right\rangle 
\left\langle g_{AB}(x;\mathbf{a,a}^\ast) \partial _{b}g_{CD}(x;\mathbf{a,a}^\ast) \right\rangle , \nonumber
\end{eqnarray}
since its real part can be \textit{exactly } rewritten as \cite{caia}%
\begin{equation}
{\rm Re}\, \widetilde{G}_{ab}(\theta )=\left\langle
\partial_{a}\,L_{AB}^{1/2}(x;\mathbf{a,a}^\ast)
\,\partial_{b}\,L_{CD}^{1/2}(x;\mathbf{a,a}^\ast) \right\rangle_{HS},
\end{equation}%
where $L_{AB}$  (non-diagonal representation) \cite{diegoFund, klauder} is given by
\begin{eqnarray}
L_{AB}& =&\int \frac{d^{2}\alpha}{\pi}\left[\int \frac{d^{2}w}{\pi }\int 
\frac{d^{2}\alpha ^{\prime }}{\pi }e^{-\left( \frac{m}{\sqrt{2}\left\vert 
\mathbf{a}\right\vert }\right)^{2}
\left[ \left(\alpha +\alpha^{\ast}\right)- B\right]^{2}+ D } e^{\xi\varrho \left(\alpha^{\prime}+\alpha^{\prime \ast }\right)}
\left\vert f\left( \xi \right) \right\vert^{2} \times\right. \\
&&\hspace{1.8cm} {\left( \begin{array}{c} \alpha^{\prime} \\ \alpha^{\ast\prime}\end{array}\right)_{\!\!\!\!(m+n)AB}}
\left. \!\! e^{\frac{\left\vert w\right\vert ^{2}}{4}}e^{\frac{i}{2}%
\left[ \left( \alpha -\alpha ^{\prime }\right) w^{\ast }+\left( \alpha
^{\ast }-\alpha ^{\ast \prime }\right) w\right] }\right] \left\vert \Psi
_{m}\left( \alpha \right) \right\rangle \left\langle \Psi _{n}\left( \alpha
^{\prime }\right) \right\vert \nonumber
\end{eqnarray}
with $m,n = 1/4, 3/4 \, (m\neq n)$, and $\left\langle \,{,}\,\right\rangle_{GS}$ 
standing for the customary inner product in the Banach space of the Gram-Schmidt operators in $\mathcal{H}$. 
The complex numbers $\alpha$ and $\alpha^\ast$ in the exponential factor are the {\em eigenvalues} of the coherent states.

From the above expression, the corresponding Gram-Schmidt operator reads
\begin{equation}
G=\int \frac{d^{2}\alpha }{\pi }\left[ \int \frac{d^{2}w}{\pi }e^{\frac{%
\left\vert w\right\vert ^{2}}{4}}e^{\frac{i}{2}\left[ \left( \alpha -\alpha
^{\prime }\right) w^{\ast }+\left( \alpha ^{\ast }-\alpha ^{\ast \prime
}\right) w\right] }\right] \left\vert \Psi _{m}\left( \alpha \right)
\right\rangle \left\langle \Psi _{n}\left( \alpha ^{\prime }\right)
\right\vert 
\end{equation}

In summary, our results come to realize the conexion conjectured by Caianiello.

\subsection{Reparameterization invariant formulation (going towards first principles)}

In information geometry, reparameterization invariance can be seen as a change of coordinates in a Riemannian manifold,
the intrinsic properties of the curvature remaining unchanged in different parameterizations. 

The Fisher metric in its original form, is not invariant under  reparameterizations. 
Why it is important to have reparameterization invariance? 
If we can establish a geometrical relation between the probability distribution, $P$,
and the metric state solution, $g_{ab}(t)$ (or the geometrical Lagrangian) via the probability current, 
it would be desirable to have reparameterization invariance with respect to the evolution parameter of the physical system,
in order to have an unique perspective of evolution of the system.
One can go further in such direction and propose the following (Nambu-Goto inspired) metric
\begin{equation}
G_{ab}=4\int {d^{D}x\,}\sqrt{\,\partial _{a}{P}^{1/2}{(x;\theta )}\,\partial _{b}{P%
}^{1/2}{(x;\theta )}}, \label{nambulike}
\end{equation}
that is full invariant under reparameterization  with respect to the physical evolution of the system.
This possibility will be explored in future work.

\section{Concluding remarks}\label{sec:conclusions}

In this work we have analyzed several aspects of the geometrical meaning of the Fisher's metric definition.
A new generalization of the Hitchin prescription for constructing the Fisher's information metric was presented, 
taking a geometrical Lagrangian as probability distribution ($P \leftrightarrow {\cal L}$).
The results were confronted with a completely different prescription:
to take as probability distribution the state probability density 
of an emergent metric (coherent) state, 
which is a solution for a non-degenerated superspace obtained in a previous work  ($P \leftrightarrow j_0$).
We then analyze the bosonic ($B_0$) part of the Fisher supermetric by putting all fermionic fields to zero,
in order to compare our solutions with those in the literature.

The main results of this research can be summarized as follows:

i) The choice of the complex constants $\mathbf{a}$ and $\mathbf{a}^\ast$
as our set of (physically meaningful) parameters is based on that they are responsible 
for the localized Gaussian behaviour of the physical states. 
This lead to a Fisher's metric on a complex 2-dimensional manifold presenting notably different behaviours in the
two approaches. In the first one ($P \leftrightarrow {\cal L}$) the Gaussian behaviour of the metric state solution $g_{ab}(t)$
was preserved while in the second one ($P \leftrightarrow j_0$) it was completely lost.
However, it is important to remark that, in both cases,
the ultralocal characteristic behavior of $g_{ab}(t)$ is preserved through a singular ($|\mathbf{a}|^{-1}$) term.

ii) In principle, it should possible to relate, from the quantum point of view, 
the $\mathbf{a}$- $\mathbf{a}^\ast$ complex manifold (Fisher's) metric with an invariant metric 
on a K\"{a}hler or on a projective Hilbert space ($CP^1$).

iii) As shown in Section \ref{sec:geom}, the function $P$, 
in sharp contrast with the Hitchin's proposal, 
can be put in direct relation with the spacetime line element $ds^{2}$ 
by making the identification $P^{1/2}\equiv \mathcal{L}_{g}$ in the same Fisher's formula. 

iv) Also in Sec. \ref{sec:geom} we demonstrate that, using the probability current $j_0$ as the probability density $P$, 
the quantum counterpart of the Fisher's metric can be exactly implemented,
and all the quantum operators involved in the geometrical correspondence, exactly constructed, 
as already inferred on a general basis by Caianiello {\em et al.} in \cite{caia}.

\section{Acknowledgments}
The authors would like to thank CNPq and PROCAD/CAPES for partial financial support.


\begin{thebibliography}{9}

\bibitem{diegoSuper} D.J. Cirilo-Lombardo, arXiv:1205.5883.

\bibitem{diegoFund} D.J. Cirilo-Lombardo, Foundations of Physics, \textbf{37} (2007) 919; \textbf{39} (2009) 373.

\bibitem{diegoPLB} D.J. Cirilo-Lombardo, Physics Letters B, \textbf{661} (2008) 186.

\bibitem{pash} D. V. Volkov and A. I. Pashnev, Theoret. and Math. Phys. {\bfseries 44}:3 (1980), 770.

\bibitem{va} V. P. Akulov, D. V. Volkov, Theoret. and Math. Phys. {\bfseries 41}:2 (1979), 939.

\bibitem{casal} R. Casalbuoni, Phys. Lett B, \textbf{62} (1976) 49.

\bibitem{fisher} R. Fisher, Math. Proc. of the Cambridge Philosophical Society {\bfseries 22} (1925) 700.

\bibitem{rao} C. R. Rao, Bull. Cal. Math. Soc. {\bfseries 37} (1945) 81.

\bibitem{frieden}  B. Roy Frieden,{\em Science from Fisher Information: A Unification},
 Cambridge Univ. Press. (2004) ISBN 0-521-00911-1.

\bibitem{plastino} L.P. Chimento, F. Pennini and A. Plastino, Phys. Lett. A {\bfseries 293} (2002) 133;
\bibitem{fp} B. Roy Frieden and A. Plastino, Phys. Lett. A {\bfseries 272} (2000) 326.

\bibitem{ponjas} U.Miyamoto and S.Yahikozawa, Phys. Rev.{\bfseries E 85} (2012) 051133 .

\bibitem{maldacena} For a comprehensive review see Ofer Aharony, Steven S. Gubser, Juan Maldacena, Hirosi Ooguri and Yaron Oz, 
Phys. Repts. {\bfseries 323} (2000) 183. 

\bibitem{hitchin} N. J. Hitchin, {\em The geometry and topology of moduli spaces}, Springer (1999).

\bibitem{caia} E.R. Caianiello, W. Guz, Phys. Lett. A {\bfseries 126}:4 (1988) 223.

\bibitem{klauder} Jhon R. Klauder and Bo-Sture K. Skagerstam,  J. Phys. A {\bfseries 40} (2007) 2093.

\end{thebibliography}
\end{document}